# Quantum Darwinism as a Darwinian process

John Campbell

## *Abstract*

The Darwinian nature of Wojciech Zurek's theory of Quantum Darwinism is evaluated against the criteria of a Darwinian process as understood within Universal Darwinism. The characteristics of a Darwinian process are developed including the consequences of accumulated adaptations resulting in adaptive systems operating in accordance with Friston's free energy principle and employing environmental simulations. Quantum theory, as developed in Zurek's research program and encapsulated by his theory of Quantum Darwinism is discussed from the view that Zurek's derivation of the measurement axioms implies that the evolution of a quantum system entangled with environmental entities is determined solely by the nature of the entangled system. There need be no further logical foundation. Quantum Darwinism is found to conform to the Darwinian paradigm in unexpected detail and is thus may be considered a theory within the framework of Universal Darwinism. With the inclusion of Quantum Darwinism within Universal Darwinism and the explanatory power of Darwinian processes extended beyond biology and the social sciences to include the creation and evolution of scientific subject matter within particle physics, atomic physics and chemistry, it is suggested that Universal Darwinism may be considered a candidate 'Theory of Everything' as anticipated by David Deutsch.

## *Introduction*

Wojciech Zurek's theory of Quantum Darwinism (Zurek, 2009)posits a Darwinian process responsible for the emergence of classical reality from its quantum substrate. This theory explains information transfer between the quantum and classical realm during the process of decoherence as involving a 'copy with selective retention' mechanism characteristic of Darwinian processes.

Quantum theory is frequently applied to microscopic systems near the limits of experimental perception. Such systems are often characterized as fundamental but this might be due more to the limitations of scientific technology than to an actual fundamental or simple nature. It is clear that our leading candidates for the next generation of fundamental physical theories will have to account for phenomena occurring at the Plank scale, that is about $10^{-35}$ meters (Smolin, 2001 ; Green 1999). When we consider that the diameter of a proton is about $10^{-14}$ meters we see that there is a greater relative difference in scale between the Plank scale and that of 'fundamental' particles than there is between that of fundamental particles and phenomena in our every day experience. There are well developed physical models that indicate quantum theory may apply at scales close to the plank scale (Markopoulou & Smolin, 2004). We might well expect there to be a good deal of emergent complexity and perhaps even layers of emergent phenomena below the scale of 'fundamental' particles. This consideration may caution us against making assumptions concerning the complexity of quantum systems.

Universal Darwinism, the collection of scientific theories that employ a Darwinian process to explain the creation and evolution of their subject matter, offers such explanations throughout a wide range of the biological and social sciences (Campbell, 2009). The question examined here is whether Quantum Darwinism might be interpreted as a theory extending the scope of Universal Darwinism into the realm of quantum phenomena: particle physics, atomic physics and chemistry amongst others.

Answering this question may be reduced to determining if the Darwinian process of Quantum Darwinism as elucidated by Zurek and colleagues is a Darwinian process as understood by Universal Darwinism.

Universal Darwinism considers a Darwinian process as a substrate-neutral algorithm that may be extracted from Darwin's theory of natural selection. This algorithm is composed of three steps: replication or copying, variations amongst the copies, and selective survival of the copies determined by which variations they possess. It has been shown that any entity embodying this algorithm will evolve and will evolve to be better at surviving. A Darwinian process operating as a system's evolutionary mechanism may often result in the system becoming an adaptive system. Adaptive systems may be characterized as having an internal model of their environment which is maintained through Bayesian updating and subject to the free energy principle (Friston, 2007). They also tend to accumulate adaptations that function through the principle of maximum entropy to place constraints on the system's evolution to states of higher entropy (Campbell , 2009).

Zurek (2009) has indicated that in his view Quantum Darwinism meets at least some of these criteria:

> *In the end one might ask: "How Darwinian is Quantum Darwinism?" Clearly, there is survival of the fittest, and fitness is defined as in natural selection.*

However he also makes it clear the exact relationship between Quantum Darwinism and natural selection is murky and requires some clarification (Zurek, 2009):

> *Is Quantum Darwinism (a process of multiplication of information about certain favored states that seem to be a "fact of quantum life") in some way behind the familiar natural selection? I cannot answer this question, but neither can I resist raising it.*

This paper is an attempt to shed some light on these questions. Section I will examine the nature of Darwinian Processes as they have come to be developed within Universal Darwinism. Section II will examine the nature of Quantum Darwinism and Section III will evaluate Quantum Darwinism as a Darwinian process. Section IV discusses some of the many interpretational issues for quantum theory raised by Zurek's work.

## I. Darwinian Processes

The algorithmic nature of natural selection allows its essential mechanism to be abstracted and hypothesized as a possible mechanism operating in the evolution of subject matter often other than biological. Numerous theories of this type abound in the social sciences and even in the hard sciences such as physics (Campbell, 2009).

The essential abstraction from natural selection, which we are calling a Darwinian process, has been developed in the work of Richard Dawkins, Daniel Dennett and Susan Blackmore (amongst others) to consist of a three-step process:
1) Replication of system.
2) Inheritance of some characteristics that have variation amongst the offspring.
3) Differential survival of the offspring according to which variable characteristics they possess.

It has been proposed that any system adhering to this three-step algorithm, regardless of its substrate, must evolve and will evolve in the direction of an increased ability to survive (Dawkins, 1976; Dennett, 1995; Blackmore, 1995).

The physical mechanisms instantiated by Darwinian processes that bestow enhanced survivability are referred to as adaptations. Adaptations are usually discovered through processes with a random 'trial and error' component, such as genetic mutation, but the greater survivability they bestow allows them to become widespread within the population. Systems with long evolutionary histories come to accumulate many adaptations. Indeed any organism may be considered as largely an accumulation of adaptations built up over evolutionary time.

Given the second law of thermodynamics, the survivability of a low entropy system is highly unlikely and must entail an intricate balancing of low entropy within the system against an increase of environmental entropy. This complex balancing requirement implies a deep 'knowledge' of the environment by the system. Theorists such as Henry Plotkin (1993) have concluded that adaptations must contain detailed information concerning their environment to the extent that they can be considered as a form of knowledge.

For this reason a Darwinian process that has been in operation for a significant period of time may become a good deal more sophisticated than the bare bones Darwinian algorithm might indicate. Take for example natural selection, the basic Darwinian process operating in the biological realm. The base 'copy with selective retention' algorithm is still in effect but has been complemented with an array of embellishments increasing its effectiveness. For instance it is a consensus amongst those who research the subject that DNA, the substance used in constructing genetic copies, is not the original substance used. At some point the use of DNA evolved and completely replaced the original mechanism. Other embellishments such as sexual reproduction have also added to the complexity with which the basic algorithm is carried out.

Such enhancements to the Darwinian process also bestow greater survivability and are thus also adaptations. Once a Darwinian process has evolved a sufficient depth of adaptations it may be considered an adaptive system. Karl Friston (2007) defines the general characteristics of an adaptive system as one that can react to changes in its environment by changing its interactions with the environment to optimize the results for itself. He examines the system in term of three features or variables: the system itself, the effect of the environment on the system and the effect of the system on the environment. These three features are integrated via an internal model that resides within the system. Friston has shown that the measures taken by adaptive systems in order to optimize their interaction with the environment can be understood via the Free Energy Principle: an adaptive system will attempt to minimize the free energy which bounds the 'surprise' resulting from its actual experience in the environment. In other words the system will act so as to minimize the prediction error of its internal model.

The internal model is constructed in a Bayesian process that makes inferences from environmental data and also utilizes knowledge of the actions which the system can take to influence outcomes in the environment. Examples of such internal models residing within adaptive systems may include genetics residing within organisms, mental models residing within brains and business plans residing within corporations.

Let's say we see something out of the corner of our eye that might be important. Because we cannot see the object very well we may have trouble assigning accurate probabilities to the numerous imaginable possibilities. Being adaptive systems we might turn our head and/or eyeballs and bring the thing into focus. We gain a pertinent data set and update, in a Bayesian manner, the probabilities associated with our internal model of the external world in conformance with this data. We change the way we interact with our environment by bringing a potentially important aspect of our environment into focus. We are now better informed and in a better position to optimize our response.

The system's interaction with its environment must be designed to optimize the outcome for itself. Thus an adaptive system must envision the optimal but realistic outcome of environmental interactions and execute those actions under its control necessary to achieve the envisioned outcome.

From this view the modeling performed by an adaptive system functions as a simulation of its environmental interactions and is thus a form of computation. The theory of computation as outlined by Turing and the Church-Turing-Deutsch principle makes the claim that there exist universal computing machines capable of processing all computable functions. David Deutsch has shown that the composition of this underlying class of computable functions is not defined by logic or mathematics but rather by the laws of nature. What is computable are those functions where nature has provided the machinery capable of performing the algorithm. In Deutsch's words (Deutsch, 1985):

> *The reason why we find it possible to construct, say, electronic calculators, and indeed why we can perform mental arithmetic, cannot be found in mathematics or logic. The reason is that the laws of physics 'happen to' permit the existence of*

> *physical models for the operations of arithmetic such as addition, subtraction and multiplication.*

Some researchers consider the computational aspect of adaptive systems as central to understanding them. Nobel laureate Sydney Brenner (1999) has called for biological science to focus on developing the theoretical capabilities required to compute organisms' phenotype from their genotype. This approach is judged to have explanatory potential as he characterizes biological systems themselves as essentially information processing machines.

It is interesting that we find computational machinery central to the functioning of adaptive systems found in nature and that it is such computational machinery found in nature that underlies the theory of computation.

Adaptive systems, by this definition, contain a good deal of internal machinery. They are complex, low entropy structures. The primary outcome of environmental interactions that any adaptive system must achieve in order for it to be an optimal encounter is to retain its low entropy structure, to survive intact. This is not a trivial accomplishment, as the second law of thermodynamics states that such outcomes are extremely rare and that their rarity increases exponentially with time and the amount of matter within the system.

An adaptive system must orchestrate encounters with its environment in such a way as to maintain its own low entropy state at the expense of increased entropy in the environment. Selecting and executing these rare outcomes requires knowledge; both knowledge of the causes operating in the environment and of how other entities existing in the environment can be exploited to maintain the system's low entropy state.

For example if we consider life forms as adaptive systems we should expect them to utilize adaptive mechanisms for lowering entropy. Sunlight is an abundant energy source at the earth's surface but only extremely specific encounters with this energy will result in work being done in the service of local entropy reduction. Life acquired knowledge to perform this feat billions of years ago when it evolved photosynthesis and this knowledge has since become ubiquitous amongst the plant kingdom. Photosynthesis causes energy in sunlight to be transferred to a high energy bond in molecules of ATP. The energy in this molecular bond is then used as a resource for performing work through complex chemical processes which in the end serve to retain the organism's low entropy state.

The knowledge required to perform photosynthesis and to use the ATP molecule is contained in the genetic code of organisms and this knowledge is passed between generations through copying of the organisms' genomes. These adaptations were created and have evolved to their present states due to the operation of the Darwinian process called natural selection.

All life forms possess these internal models in the form of genetic molecules, which are central to their functioning as adaptive systems.

The principle of Maximum Entropy requires systems to move to states of highest entropy consistent with the constraints operating on the system. The ability of adaptive systems to maintain their low entropy state must therefore be explainable in terms of the constraints they deploy against increasing entropy. These constraints are precisely the adaptations developed during the evolution of the system. In our example of the production and consumption of ATP molecules within organisms those constraints often take the form of enzymes which constrain the system's chemical pathways to follow a specified course. These enzymes are directly coded for in the organism's DNA. Thus the knowledge implicit with their entropy-reducing abilities is contained in the organisms' internal models.

At the heart of Bayesian probability is the notion of updating: our beliefs in a hypothesis should be updated whenever pertinent new data becomes available. Bayes' Theorem provides the exact measure by which our confidence in a hypothesis should change with a given piece of new data. The essence of the theorem is the common sense notion that our confidence should be adjusted in accordance with the extent to which the hypothesis or model enhances our ability to predict the new data.

This theorem has been shown to be the unique mathematical description of how knowledge may be accumulated (Jaynes, 1985). It is a signature characteristic of a Darwinian process and is a central mechanism in an adaptive system's ability to model its situation (Campbell, 2009). A consequence of the importance of Bayesian updating is that we see evidence of its operation wherever knowledge accumulates in nature, whether in genetics, brain functioning or science.

In the course of this analysis of Darwinian processes we have developed a detailed a set of criteria against which any prospective Darwinian process, such as Quantum Darwinism, may be compared. These criteria include:
1. The operation of a three step 'copy with selective retention' algorithm.
2. Operation as an adaptive system where the system, the effect of the environment on the system and the effect of the system on the environment are modeled internally and the system conforms to Friston's free energy principle. The simulation of the system's environmental interactions may be understood as computation.
3. Maintenance of the internal model's accuracy through Bayesian updating.
4. Production of adaptations which utilize the principle of maximum entropy to maintain the system's low entropy state.

## *II. Quantum Darwinism*
The research program into the nature of quantum systems conducted by Wojciech Zurek of Los Alamos National Laboratory and colleagues has spanned more than 25 years and has identified and developed a number of novel processes and constructs. The most important of these might include: decoherence, einselection, envariance and the theory of Quantum Darwinism. The theoretical advances gained by this program have enabled

Zurek to derive what have commonly been considered two axioms of quantum theory from the remaining three.

Quantum theory is often presented as a derivation from a set of axioms. A common set of axioms are:
1) The state of a quantum system is represented by a vector in its Hilbert space.
2) Evolutions are unitary (i.e., generated by Schrodinger equation).
3) Immediate repetition of a measurement yields the same outcome.
4) Measurement outcome is one of the orthonormal states, the eigenstates of the measured observable.
5) The probability $p_k$ of finding an outcome $|s_k\rangle$ in a measurement of a quantum system that was previously prepared in the state $|\psi\rangle$ is given by $|\langle s_k|\psi\rangle|^2$.

Decoherence occurs when information is copied from a quantum system to its environment. This process, often referred to as a measurement, has proven an unresolved interpretational quandary for quantum theory since the Bohr-Einstein debates occurring during the first part of the last century and has become known as the measurement problem. The problem arises as many consider the axioms to contain seeming contradictions; axiom 2 requires a quantum system to evolve via the Schrodinger equation in a mathematically smooth, deterministic and continuous manner but axiom 4 and axiom 5 require the quantum system to jump in a discontinuous manner, upon measurement, to a specific state, one often much different from the state to which it had evolved prior to the measurement.

Zurek's resolution is to show that the two problem axioms are implied by the other three (Zurek, 2009). The proof of this derivation reveals that much of the information contained in the state vector after a period of unitary evolution cannot survive its being copied into the environment. Only a subset can survive the transfer. Axiom 3 requires that the information that survives in the systems state vector after the transfer is consistent with that copied to the environment resulting in a seemingly discontinuous change in the state vector.

Theoretically decoherence has been understood as consisting of two processes roughly equivalent to axioms 4 and 5 (Zurek, 2009). The first process is named 'environment induced superselection' or einselection. When a quantum system becomes entangled with an entity in its environment it is properties of the environmental entity that determine the type of information which can be copied from the quantum system into that environment. For instance some environments may be structured to receive information on position and others to receive information on momentum. Mathematically the subset of information that can survive transfer to an environment is restricted to pointer states of the system. These pointer states are the eigenstates of axiom 4 and may be predicted as they are those which will leave the system in the lowest entropy state available (Zurek, 2009). Conversely the ability of the environment to receive information from the quantum system is dependent on its ability to increase entropy (Zwolak, Quan, & Zurek, 2009).

The second major simplification of quantum theory produced by Zurek's program was the derivation of axiom 5 using symmetry properties of the entangled quantum system and environmental entities. This process, environment–assisted invariance or envariance, provides a probabilistic prediction of the measurement value.

Importantly both of these simplifications provide predictions regarding measurement; they predict the type of information that will be measured and the value of the measurement. Given Zurek's demonstration that, in quantum processes, information is physical and that there is no information without representation (Zurek, 1998 ) and that axiom 1 posits that the state of the quantum system is represented by the Hilbert state vector, we are compelled to conclude that the quantum system must contain a physical representation of its state vector. Crucially, this physical representation may be considered an internal model residing within the quantum system and this model, through the measurement predictions it contains, simulates the quantum system's interactions with its environment.

Recently Zurek's research has focused on the fate of quantum information that has been copied into its environment. His major finding is that the information best able to survive the process of decoherence is also the information having the highest reproductive success in the environment, thereby causing it to become the most widespread. This principle, along with decoherence, is encapsulated by the theory of Quantum Darwinism.

## III. Quantum Darwinism as a Darwinian process

Given these brief summaries of Darwinian processes and of Quantum Darwinism we are in a position to evaluate the extent to which Quantum Darwinism might be considered a Darwinian process. Clearly, in Zurek's view, there is little question of the Darwinian nature of quantum processes or of its central importance; he sees Quantum Darwinism as conforming to the Darwinian paradigm and identifies it as the mechanism responsible for the emergence of classical reality from the quantum substrate: (Zurek, 2004)

> *The aim of this paper is to show that the emergence of the classical reality can be viewed as a result of the emergence of the preferred states from within the quantum substrate thorough the Darwinian paradigm, once the survival of the fittest quantum states and selective proliferation of the information about them are properly taken into account.*

The Darwinian paradigm, as defined within Universal Darwinism, encompasses details in addition to survival of the fittest and selective proliferation. The following evaluation will consist of a comparison of four characteristics of Darwinian processes (listed at the end of section 1) with the characteristics of quantum systems as described by Quantum Darwinism.

1. **The operation of a three step 'copy with selective retention' algorithm.**

Decoherence essentially describes a process by which information is copied or transferred from a quantum system to its environment. Much of the varied information contained in the state vector is copied but most has extremely short periods of survival. The fittest quantum states, or pointer states, are selected (Zurek, 1998 ). The selected quantum states are those capable of the greatest reproductive success (Zurek, 2004). Clearly Quantum Darwinism is formulated in a manner consistent with the three step algorithm of a Darwinian process.

2. **Operation as an adaptive system where the system, the effect of the environment on the system and the effect of the system on the environment are modeled internally and the system conforms to Friston's free energy principle. The simulation of the system's environmental interactions may be understood as computation.**

Axiom 1 of quantum theory tells us that all information concerning a quantum system is contained within the state vector making it an excellent candidate for an internal model. As there can be no information without representation this information must physically reside within the quantum system. Axiom 2 tells us that this state vector, and the information it contains, will evolve according to the Schrödinger equation:

$$i\hbar \frac{\partial}{\partial t} \Psi(x, t) = \hat{H} \Psi(x, t)$$

Where ψ(x, t) is the state vector and $\hat{H}$ is the Hamiltonian operator.

As the Hamiltonian operator encapsulates internal effects of the system, the effect of the system on the environment and the effects of the environment on the system this model of the quantum system meets the criteria of an internal model operating within an adaptive system.

Further, we can consider Friston's free energy principle to be satisfied by this adaptive system if we note that Zurek has shown the measurement predictions of quantum theory are implied by the first three axioms. As the predictions of quantum theory are amongst the most accurate in science we might conclude that the prediction error or free energy between the model and the system's actual experiences is minimized.

In Seth Lloyd's (2007) treatment of quantum computation the interactions of quantum systems are interpreted as quantum computations. The result of the computation is the outcome of the interaction. In this sense the state vector precisely predicts the outcome and the prediction error is zero. We note in this context that the noiseless subsystems by which a quantum computation comes to its answer have been shown to be mathematically isomorphic with Zurek's process of einselection (Blume-Kohout, 2008). Given the isomorphism between these two processes we are entitled to view quantum computation as a process where the 'answer' to a computation emerges via a Darwinian process.

Given Lloyd's interpretation that outcomes of quantum interactions are equivalent to simulations carried out by the quantum system and the fact that all physical interactions are quantum interactions we are led to the Church-Turing-Deutsch principle which states that every finitely realizable physical system can be perfectly simulated by a quantum computation.

A quantum system, as described by Zurek's theory, meets the criteria of an adaptive system. The internal model formed by a representation of the state vector integrates information concerning the system, the effect of the environment on the system and the effect of the system on the environment. It also operates in accordance with Friston's free energy principle for adaptive systems. The state vector, considered as an internal model, may be interpreted as performing computational simulations of the system's environmental interactions.

3. **Maintenance of the internal model's accuracy through Bayesian updating.**

 In order for an internal model of an adaptive system to reduce its prediction error it must accumulate knowledge through inference from environmental data. It has been demonstrated that there is a single method of conducting inference, and that is Bayes' Theorem, derived from the basic sum and product rule of Bayesian probability (Jaynes, 1985). Bayes' Theorem provides a method of calculating how confidence in a hypothesis or a model should be updated when new data becomes available:

$$\mathbf{P(H|DX) = P(H|X)\frac{P(D|HX)}{P(D|X)}}$$

Where we are considering the Probability of Hypothesis (H) given prior data (X) and new data (D)

In other words we should update our probability for the hypothesis or model being true by the extent to which the hypothesis enhances our ability to predict the new data.

What was formerly axiom 4 of quantum theory reveals that every possible measurable state of the quantum system is predicted by the system state. The correct weighting that should be given to each possible outcome is given by Born's rule. With the occurrence of a measurement of the system the state vector is updated. The probabilistic weightings for the various outcome hypotheses are reassigned in a Bayesian manner in accordance with axiom 3 and the hypothesis associated with a repeat of the previous measurement result is assigned probability one and all other hypothesis are assigned probability zero. In other words our confidence as to the state of a quantum system immediately upon receiving pertinent data is strengthened to certainty.

We can conclude that the internal model of quantum systems undergoes Bayesian updating through the system's interactions with its environment.

4. **Production of adaptations which utilize the principle of maximum entropy to maintain the system's low entropy state.**

   Zurek's program has shown that the interaction outcomes involving quantum systems can be predicted by a method named 'predictability sieve' (Zurek, 2003). This method essentially ranks all potential outcomes of the state vector according to their entropy production. Those outcomes that should be predicted (in accordance with former axiom 4) are those having the lowest entropy. This ability of a system to find rare low entropy states which are also reproductively successful is a hallmark of Darwinian processes.

   As prevalence towards decoherence is strongly influenced by mass, relatively massive quantum systems such as those considered by atomic physics and chemistry are almost continuously decohered by their environments and their existence assumes what Zurek describes as a classical trajectory. This trajectory is characterized by stability, predictability and reproductive success, and should be understood as a near continuous 'copy with selective retention' process which unremittingly probes potential environmental interactions for low entropy solutions and instantiates those solutions once found. The system's low entropy state, maintained by these interactions is balanced by an increase in environmental entropy (Zwolak, Quan, & Zurek, 2009).

   The evolution of these intricate quantum systems, which may be composed of many 'fundamental' particles, is governed by a Hamiltonian operator which models the complex interactions between the system and its environment. The principle of maximum entropy requires that the ability of these low entropy structures to have a well advertised and persistent presence within their classical environments be explained in terms of the system's ability to impose constraints on the increase of entropy. These constraints might be considered adaptations and appear as scientific laws including those governing atomic orbitals and chemical bonds.

   Quantum systems evolve by continuously probing their environments to discover and instantiate rare low entropy outcomes at the expense of increasing environmental entropy. The variations adopted by the quantum system in its evolution may be considered adaptations and function to provide constraints against increasing entropy. Much of the subject matter of atomic physics and chemistry can be considered adaptations discovered and instantiated through a Darwinian process.

In this perhaps radical view quantum systems emerge into classical reality and evolve there through a succession of information transfers to the environment. Each successive generation of offspring information contain variations. The variations which survive and are accumulated over generations are those that succeed in minimizing system entropy. In a suitable environment a quantum system may accumulate atomic and molecular adaptations and function as a relatively complex composite system. Adaptive systems are characterized by internal models which simulate and orchestrate their environmental interactions. Quantum theory tells us that quantum systems must include a physical implementation of an information processing model equivalent to the evolution of a state vector in Hilbert space. While the nature of this model remains outside of experimental verification we are left in a similar situation to biology prior to 1953 when many characteristics of life's internal model were known and predictions could be made by invoking laws such as Mendelian inheritance but the manner of the model's implementation in molecular DNA was not known.

This evaluation indicates that quantum systems as described by Quantum Darwinism meet the listed criteria and may well be considered Darwinian processes as defined within Universal Darwinism.

## *IV. Interpretational implications of Quantum Darwinism*

Zurek's research program, encapsulated by the theory of Quantum Darwinism, offers a detailed explanation of the relationship between classical and quantum reality. In the author's opinion it provides the greatest theoretical improvement in our understanding of this relationship since Bohr's development of the correspondence principle in 1920. Einstein engaged Bohr in a series of debates claiming that quantum theory must be considered incomplete as long as it was unable to explain more clearly the nature of ontological reality, which for Einstein was classical reality. Philosopher C.P. Snow described these debates: 'No more profound intellectual debate has ever been conducted' (Isaacson, 2008). We might judge the importance of Zurek's program by considering that the explanations it offers would likely have satisfied both Bohr and Einstein.

The major interpretational issues for quantum theory raised by Quantum Darwinism result from its simplification and clarification of the subject through reducing the number of axioms required for its foundation. This radical reshaping of the foundations of quantum theory leaves most predictions intact but provides a much more thorough explanation of the processes involved.

Axiom 4 of quantum theory has now been explained as due to the process of einselection. In the course of developing this explanation a number of further important mechanisms have been identified. One of these is 'predictability sieve' which clarifies the ability of quantum systems to probe their environment for low entropy states which have potential existence and to instantiate those states once found.

Axiom 5 of quantum theory has now been derived through the process of envariance, which produces Born's rule. Crucially, taken together, einselection and envariance demonstrate that the measurement predictions of quantum theory are implied by the remaining three axioms of the theory. Zurek's principle of 'no information without representation' leads us to conclude that sufficient information for the predictions of quantum theory resides in some physical form within the quantum system.

The emergence of classical reality from the quantum substrate, the topic often most fully informed by two seemingly arbitrary axioms, has been explained in detail.

Hopefully this treatment will finally lay to rest the interpretational confusion around the role of a human observer in quantum measurements that has been prevalent in many treatments and taken to anthropomorphic extremes by some such as Wigner (1970). Zurek's work makes it clear that decoherence takes place whenever there is an information transfer to the environment (Zurek, 2009). No human observer need be in attendance.

The word measurement currently contains an anthropomorphic component but it is also much more nimble in usage than are many of the alternatives such as 'monitoring by the environment'. The many variants or affixes of the root 'measure' such as measurement, measurable, measuring etc. make the word attractive for use in explanations. A possible resolution might be to create an additional scientific definition of measurement based on decoherence and devoid of references to a human observer.

## *Conclusion*

Given the evaluation that Quantum Darwinism is a theory utilizing a Darwinian process and thus a component of Universal Darwinism we find Universal Darwinism extended beyond theories in biology and the social sciences to contain explanatory theories for particle physics, atomic physics and chemistry. Universal Darwinism may thus be seen as a paradigm offering a single comprehensive explanation of much of scientific reality.

David Deutsch's ground breaking work The Fabric of Reality (Deutsch, 1997) predicted that science would soon produce 'a theory of all subjects: a Theory of Everything'. This theory would provide a single paradigm explaining scientific subject matter across all subjects. He ventured to provide details concerning four explanatory strands which the theory might integrate:
1) Quantum Physics
2) Epistemology
3) The theory of computation
4) The theory of Darwinian evolution

I suggest that this discussion has shown the 1st and 4th of these strands to be integrated within the theory of Universal Darwinism.

Epistemology referred to by Deutsch as the 2nd strand is the philosophy of science. Evolutionary epistemology, a branch of philosophy within the philosophy of science, contains a number of well-established theories that ascribe the evolution of science to the operation of Darwinian processes (D. Campbell 1965; J. Campbell, 2009; Popper, 1972) and thus Deutsch's 2nd strand is integrated within Universal Darwinism.

Deutsch has argued that computability is a property bestowed by the natural world not by logic or mathematics, and that those functions that are computable are those for which nature has designed the appropriate computational machinery. The most basic computational machinery so far found in nature is that allowing quantum computation. This paper makes arguments that quantum computation may be viewed as the simulations of an adaptive system and that the 'answers' to a quantum computation emerge through a Darwinian process. These are arguments that Deutsch's 3rd strand is integrated within Universal Darwinism.

As all physical interactions are quantum interactions we might expect quantum computations to be utilized within other adaptive systems. Some evidence is accumulating in support of this notion. It has been shown that photosynthesis may utilize quantum computations to calculate the most efficient chemical pathways for the conversion of energy contained within a specific photon (Fleming et al., 2007).

The theory of Universal Darwinism, as it integrates Deutsch's four strands within a single explanatory paradigm, may serve as a candidate for the 'Theory of Everything' as anticipated by Deutsch.